\documentclass[12pt]{article}
\pdfoutput=1

\usepackage{color}
\usepackage{amsmath}
\usepackage{amsfonts}
\usepackage{amssymb}
\usepackage{graphicx}
\usepackage{subfig}             

\usepackage[english]{babel}
\usepackage[applemac]{inputenc}
\usepackage{braket}

\usepackage{cite}

\usepackage[	colorlinks=true,	citecolor=black,	linkcolor=black,	urlcolor=blue,hypertexnames=false]{hyperref}

\def\beq{\begin{eqnarray}}
\def\eeq{\end{eqnarray}}
\def\bean{\begin{equation*}}
\def\eean{\end{equation*}}
\newcommand{\arXiv}[2]{\href{http://arxiv.org/pdf/#1}{{\tt #2/#1}}}
\newcommand{\arXivold}[1]{\href{http://arxiv.org/pdf/#1}{{\tt #1}}}

\numberwithin{equation}{section} 

\newcommand{\RN}[1]{%
  \textup{\uppercase\expandafter{\romannumeral#1}}%
}

\newcommand{\bsp}{\begin{split}}
\newcommand{\esp}{\end{split}}

\newcommand{\gcmic}{\,\mathrm{g\,cm^{-3}}}
\newcommand{\erg}{\,\mathrm{erg}}

\newcommand{\eq}[1]{\begin{align}#1\end{align}}

\newcommand{\MeV}{\,\mathrm{MeV}}

\newcommand{\TeV}{\,\mathrm{TeV}}

\newcommand{\s}{\sigma}

\newcommand{\order}[1]{O(#1)}

\newcommand{\nn}{\nonumber}

\newcommand{\f}[2]{\frac{#1}{#2}}

\def\gN{\left(\frac{C_{N}}{f_a}\right)}
\def\Gs{\Gamma_\s}
\def\l{\lambda}
\def\intinfzero{\int^{\infty}_0}
\def\gmgm{\gamma\gamma}

\begin{document}
\begin{titlepage}

\begin{center}

	{	
		\Huge \bf 
		Revisiting Supernova 1987A Limits}\vskip.15cm{\Huge \bf on Axion-Like-Particles
	}
	
\end{center}
	\vskip .3cm
	
	\renewcommand*{\thefootnote}{\fnsymbol{footnote}}


\begin{center} 
{\bf \
Jun Seok Lee\footnote{\tt \scriptsize
		 \href{mailto:phylee@ucdavis.edu}{phylee@ucdavis.edu}
		 }

		} 
\end{center}

	\renewcommand{\thefootnote}{\arabic{footnote}}
	\setcounter{footnote}{0}


\begin{center} 

	{\it Department of Physics, University of California Davis\\One Shields Ave., Davis, CA 95616}

\end{center}


\centerline{\large\bf Abstract}

\begin{quote}
Using recent supernova models, I revisit the Supernova 1987a constraints on scalar/pseudoscalar Axion-Like-Particles (ALPs). On the basis of the neutrino detections, the luminosity of ALPs must be $\lesssim 5\times10^{52} \erg/s$ after the core-bounce, and this bound limits ALPs couplings. Contrary to the QCD axion models where all couplings are $\sim 1/f_a$, it has been shown recently that a radion/dilaton could have far different coupling strength to photons and to nucleons. This fact has raised a need to establish the limits on each coupling independently. The bounds from ALPs emission by nucleon-nucleon bremsstrahlung through a two-nucleon coupling and by the Primakoff process through a two-photon coupling are updated considering the total volume emission. I find the bounds on the two-photon coupling from the Primakoff process differ from previous bounds by an order of magnitude. Through the volume emission study, trapping regimes for $m_a\gtrsim10\MeV$ are also alleviated, which need to be probed by future experiments.
\end{quote}

\end{titlepage}


\section{Introduction}

Astrophysical objects provide a powerful natural laboratory in elementary particle physics, and stars are the best sources of weakly interacting particles such as neutrinos, gravitons, radions, and axions. Supernova 1987a (SN1987a) is one of the most important astrophysical sources due to its high density, high temperature, and proximity.

A massive progenitor star eventually goes through core-collapse and supernova explosion (a type \RN{2} scenario) where the bouncing shock wave ejects the entire overburden of the mantle and the envelope. It radiates its gravitational binding energy in the form of neutrinos within a few seconds. Since the neutrino detection on February in 1987 coming from SN1987a, it has been widely used to constrain exotic light particles because an extra degree of freedom carrying off the energy from the star will affect the cooling time of the star, hence the measured neutrino detection duration. Despite the fact that it was only 25 neutrinos detected by IMB~\cite{Bionta:1987qt}, Kamiokande \RN{2} ~\cite{Hirata:1987hu} and Baksan~\cite{Alekseev:1987ej} lasting less than 13 seconds, the neutrino burst observation provides essential constraints on exotic light particles.

Many beyond the Standard Model (BSM) physics scenarios include new light (pseudo-)scalars in their spectra. One example of a BSM pseudoscalar is the QCD axion, a dynamical solution to the strong CP problem~\cite{Peccei:1977hh,Wilczek:1977pj,Weinberg:1977ma}. One example of a BSM scalar is the radion which arises from introducing an extra dimension~\cite{LightRadion,Bellazzini:2012vz,Csaki:2007ns,Csaki:2000zn, Bellazzini:2013fga,Chacko:2013dra,Csaki:1999mp,Csaki:1999jh,Chacko:2014pqa}. In many scenarios, their coupling and mass are not independent. The Axion-Like-Particles (ALPs) are generic parametrization of exotic light (pseudo)scalars where one treats its coupling independently of its mass. Throughout this paper, ALPs will denote any (pseudo)scalars which interact with two photons or nucleons.

While the strongest bound from SN1987a is from the nucleon coupling of ALPs due to its high baryon density, this bound is usually directly translated to the two-photon coupling of ALPs. This is because this dimensionful coupling shares the same energy scale $f_a$, as the $U(1)_{PQ}$ breaking scale for the axion for example, and the dimensionless coefficients for each operator are naturally chosen to be $\order 1$. However, it has been shown that this is not the case for a dilaton or a radion~\cite{LightRadion,Bellazzini:2012vz}. In that case, even though each dilaton/radion operator has a universal scale-invariance-breaking scale, the coupling coefficients are independent of each other depending on the $\beta$-function coefficient. Furthermore, each coupling constraint can be alleviated or even avoided with the aid of the other~\cite{LightRadion}. This fact means that the constraint plots on $f_a$-$m_a$ parameter space without carefully stating which operator the constraints are imported from can be misleading and raises a need to establish the limits on each operator independently. As supernovae have both very high baryon density as well as high plasma density, the SN1987a constraints on the coupling to nucleons and on the coupling to photons has to be investigated separately keeping in mind that having one of the couplings in the trapping regime, the essential feature of SN bound, will make the other coupling constraint-free~\cite{LightRadion}.

The goals of this paper are to update the ALP constraints with a more recent supernova simulation~\cite{Fischer} and to discuss limits on the nucleon coupling and the photon coupling individually. It will give readers a detailed review of the well-known SN1987a limits and a guide for how one should read SN1987a limits on the Axion-Like-Particle parameter space.

A brief outline of the paper is as follows.
I review ingredients of SN1987a bounds in Section~\ref{PriorSN}, and provide the supernova models I use for this work in Section~\ref{SNmodels}. I calculate the ALP production rate and absorption rate by nucleon-nucleon bremsstrahlung and Primakoff process in Section~\ref{ProdAbsp}. Special attention is given to what assumptions and corrections are made for uncertainties inevitably present for supernova models. The main result plots are shown in Section~\ref{Chap:Results}. I close out by presenting some brief conclusions in Section~\ref{Chap:Conclusions}.

\section{Prior Supernova 1987A Limits on ALP}\label{PriorSN}
\setcounter{equation}{0}

\begin{figure}[t]
 \centering
  \includegraphics[height=.45\textheight]{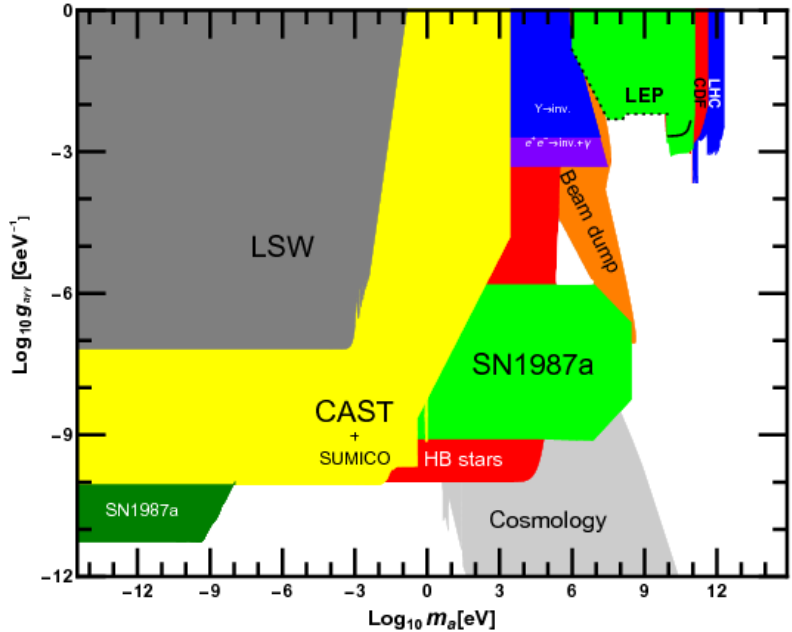}
  \caption{Limits on ALP coupling and mass parameters space compiled by Jaeckel, Jankowiak and Spannowsky~\cite{Jaeckel:2015jla} and the references therein.}
  \label{fig:ALPConstraints}
\end{figure}

The measurements of the neutrino burst from Supernova 1987a set important constraints on sub-GeV ALPs. An approximate analytic constraint on the energy loss for SN1987a is set by the neutrino burst duration~\cite{Burrows:1988ba} detected by IMB, Kamiokande \RN{2} and Baksan. From the measured cooling rate, the energy loss rate due to beyond-the-standard-model particles should not exceed the energy loss rate through neutrinos~\cite{Raffelt:1996wa}:
\beq\label{totEbound}
\dot{E}_{\mathrm{new}} \lesssim 5\times10^{52} \erg/s,
\eeq
given the fact that the neutrino luminosity in all six degrees of freedom without any exotic degree of freedom is compatible with the measurements. This criterion has been considered by many authors to constrain the exotic light particles. The light green region in Fig.~\ref{fig:ALPConstraints} shows a constraint on the coupling to photons from SN1987a when other couplings are all neglected~\cite{LightRadion,Masso:1995tw,Jaeckel:2015jla}. This excluded limit covers an ALP mass near 30 MeV or less, with a coupling to photons suppressed by a scale between $10^3\TeV$ and $10^6 \TeV$.

One exceptional feature of the SN1987a bound is that the limit has the upper bound. This is because supernovae are so dense that in the strong coupling regime, the star is so opaque to the radiating exotic particles and traps the particles before they escape. This regime is called the trapping regime, which is free from supernova bounds. This trapping regime is significant in that while ALP mass smaller than 100 keV is highly constrained from other constraints as shown in Fig.~\ref{fig:ALPConstraints}, the trapping by nucleons can relieve the SN1987a constraints on the photon coupling or vice versa~\cite{LightRadion}.

Since there are many aspects of SN1987a that are poorly understood, the free streaming regime has been set roughly by taking the energy loss rate per volume,
\beq\label{volEbound}
\dot{\varepsilon}_{\mathrm{new}} \lesssim 3\times10^{33} \erg\, \mathrm{cm}^{-3}\,s^{-1}.
\eeq
to be calculated at the typical core density of around $3\times10^{14} \gcmic$ and temperature of around $30$ MeV. This gives a good estimated bound in the free streaming regime. In the trapping regime, things are more complicated as the particles will be deposited at different locations and will transfer energy from core to mantle in the same way as radiative transport. For good treatment of trapping regime, one needs a better understanding of the supernovae structure.


%


\section{SN1987a Models}\label{SNmodels}
\setcounter{equation}{0}

\begin{figure}[thpb]
 \centering
  \includegraphics[height=.29\textheight]{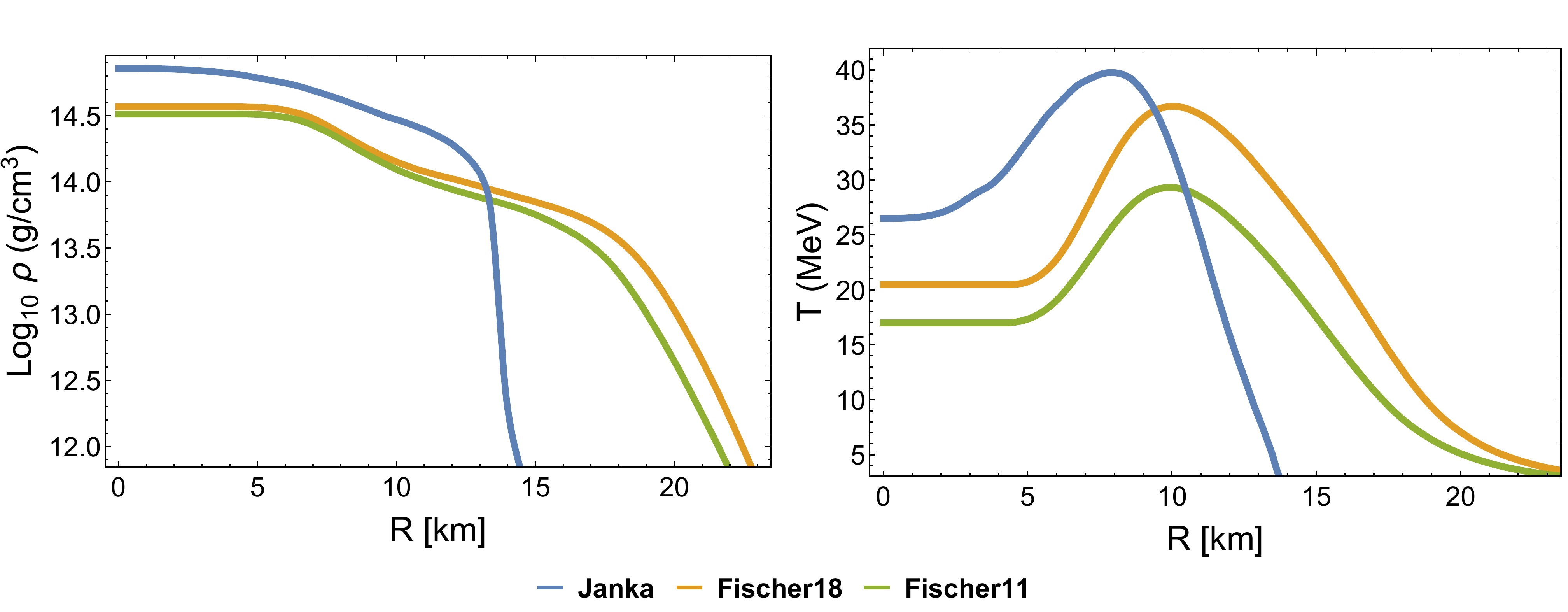}
  \caption{Profile functions of density and temperature for the three models considered in the text.}
  \label{Profiles}
\end{figure}

For this work, I'll impose the two numerical SN models studied in the paper by Fischer et al.~\cite{Fischer}. They employ a spherically symmetric core-collapse SN model AGILE-BOLTZTRAN for a precollapse progenitor mass of 18.0 M$_{\odot}$ and 11.2 M$_{\odot}$, which I'll refer ``Fischer18" and ``Fischer11" respectively. Along with these models, I also present one of the most popularly used protoneutron star model, S2BH0 in~\cite{Keil:1994sm} by Janka et al., as a fiducial model to be referred as ``Janka." The profile functions of density and temperature 1s after the core bounce are plotted in Fig.~\ref{Profiles}. 


The three models, Fischer18, Fischer11, and Janka, have different total masses. Respectively they are: $1.43\, M_\odot, 1.14\, M_\odot,$ and $1.53\, M_\odot$. The effective radii are: 33.8 km, 34.6 km, and 15 km.

\section{ALP Production and Absorption}\label{ProdAbsp}
\setcounter{equation}{0}
\subsection{N-N Bremsstrahlung}\label{Bremsstrahlung}
\begin{figure}[thpb]
 \centering
  \includegraphics[height=.15\textheight]{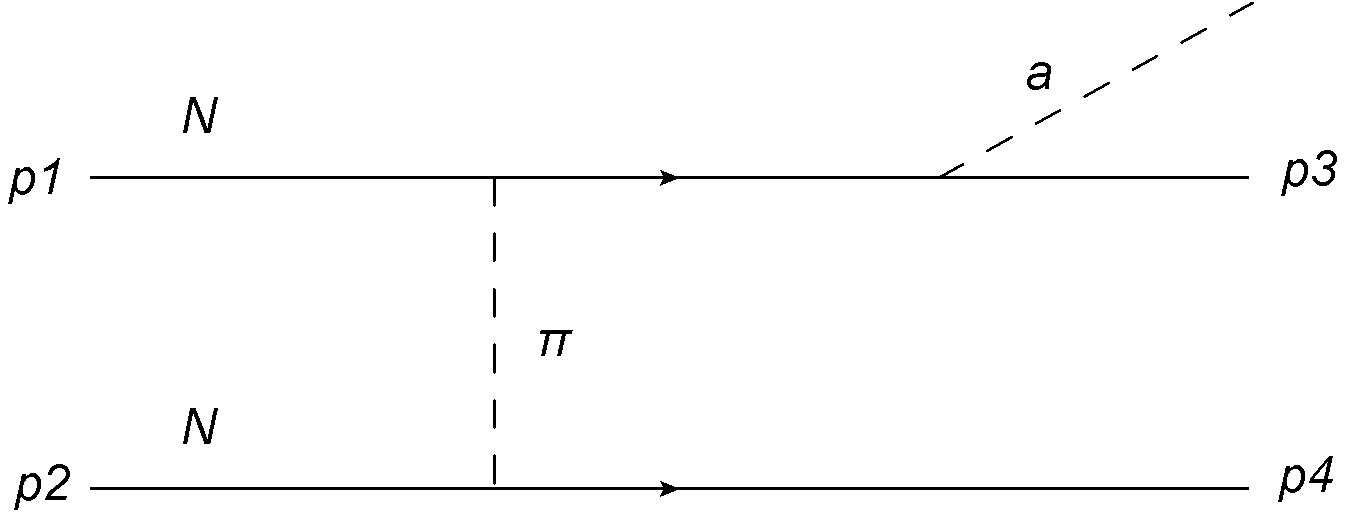}
  \caption{Nucleon-nucleon bremsstrahlung with one pion exchange.}
  \label{fig:NNbrem}
\end{figure}

The ALPs are produced by nucleon-nucleon bremsstrahlung through the one pion exchange (OPE) process. One of the eight diagrams is shown in Fig.~\ref{fig:NNbrem}.

We will assume that the matter in the core of SN1987a is mostly non-relativistic nucleons, i.e., $T\ll 1$ GeV. The energy loss rate per unit volume from nucleon-nucleon bremsstrahlung ($NN\rightarrow NNa$) and the absorption rate of ALP having energy $w$ in the nucleon medium ($NNa\rightarrow NN$) are given by phase space integrals of the squared amplitudes
\begin{align}
\dot{\epsilon}&=\int d\Pi_1d\Pi_2d\Pi_3d\Pi_4d\Pi_\sigma (2\pi)^4 \delta^4(p_1+p_2-p_3-p_4-p_\sigma)E_\sigma      \nn\\
            &\qquad \times \frac{1}{4}\sum_{\mathrm{spins}}|\mathcal{M}|^2f_1f_2(1-f_3)(1-f_4), \label{epdotint}
\end{align}
\begin{align}
\Gamma_w     &= \frac{1}{2w}\int d\Pi_1d\Pi_2d\Pi_3d\Pi_4 (2\pi)^4 \delta^4(p_\sigma+p_1+p_2-p_3-p_4)     \nn\\
        &\qquad \qquad \times \frac{1}{4}\sum_{\mathrm{spins}}|\mathcal{M}|^2f_1f_2(1-f_3)(1-f_4), \label{lambdainvint}
\end{align}
where $p_1,\, p_2,\, p_3,\, p_4,$ and $p_\sigma$ are the nucleon and radion four-momenta with the subscripts 1 and 2 (3 and 4) for the incoming (outgoing) nucleons;
\beq
d\Pi_i=d^3p_i/(2\pi)^32E_i
\eeq
 is the Lorentz invariant phase-space volume element; $f_i$ are the nucleon phase-space distribution functions.


The estimate for $\dot{\epsilon}$ in (\ref{epdotint}) assumes that the mean free path is much larger than the size of the region of high nuclear density. When the mean free path becomes smaller than this size, some of the radions produced in the SN will be absorbed before escaping. We can give an improved estimate of the energy loss rate due to ALPs that takes into account the ALPs absorption by the following formula
\beq
\label{Eloss}
\dot{E}=\int^{R}_{0}4\pi r^2 dr~\dot{\epsilon}(T,\rho)\exp\left[\int^{\infty}_r \f{dl} {\lambda(T,\rho)}\right]~\mathrm{erg~s^{-1}},\label{totaledot}
\eeq
where $R$ is the effective radius of the star 1s after the core bounce where the density and temperature becomes negligible compared to the core, $\dot{\epsilon}(T,\rho)$ is the energy loss rate per volume \eqref{epdotint}, and $\lambda(T,\rho)$ is the reduced Rosseland mean free path obtained using \eqref{lambdainvint}. The profile functions of temperature and mass density for SN1987a are given in Sec.~\ref{SNmodels}



\subsubsection*{Pseudoscalar Matrix Element}\label{Pseudo}

A pseudoscalar interaction with nucleons is parametrized as
\beq
\mathcal{L}_{aNN}=\frac{C_{aN}}{2f_a}\bar{\psi}_N\gamma_\mu \gamma_5 \psi_N \partial^\mu a,\label{psL}
\eeq
where $f_a$ is an energy scale (i.e. Peccei-Quinn symmetry breaking scale for axions), $C_{aN}$ with $N=n$ or $p$ is a dimensionless coupling parameter and a factor of one half is conventional for a derivatively coupled interaction form which will be cancelled in equation of motion when we integrate by parts. 
The squared amplitude of the OPE diagrams for a pseudoscalar (neglecting $\eta$ meson exchange channels) is found to be~\cite{Raffelt:1996wa}:
\begin{align}
\sum_{\mathrm{spins}}|\mathcal{M}|=&\frac{256}{3}\left(\frac{C_{aN}}{f_a}\right)^2\frac{f^4 m_N^4}{m_\pi^4}\left[\left(\frac{\mathbf{k}^2}{\mathbf{k}^2+m_\pi^2}\right)^2+\left(\frac{\mathbf{l}^2}{\mathbf{l}^2+m_\pi^2}\right)^2+\frac{\mathbf{k}^2\mathbf{l}^2-3(\mathbf{k}\cdot\mathbf{l})^2}{(\mathbf{k}^2+m_\pi^2)(\mathbf{l}^2+m_\pi^2)}\right],\label{psME}
\end{align}
where $\mathbf{k}=\mathbf{p_2-p_4}$, $\mathbf{l}=\mathbf{p_2-p_3}$, and $f\approx1$ is the pion-nucleon dimensionless coupling. The square bracket is $\order{1}$ in the SN thermal medium since $|\mathbf{k}|$, $|\mathbf{l}|\sim\sqrt{3m_N T}\gtrsim m_\pi$ so that we can approximate the matrix element as a constant:
\begin{align}
\sum_{\mathrm{spins}}|\mathcal{M}|\simeq&\frac{16}{3}g_{aNN}^2g_{\pi NN}^4,\label{psM}
\end{align}
where $g_{aNN}\equiv\frac{C_{aN}}{f_a}$ and $g_{\pi NN}=\frac{2 f^4 m_N^4}{m_\pi^4}\approx13$ phenomenologically.


\subsubsection*{Scalar Matrix Element}\label{Scalar}

A scalar interaction with nucleons are parametrized as
\beq
\mathcal{L}_{s NN}=\frac{C_{s N}}{2f_s}\bar{\psi}_N\gamma_\mu \psi_N \partial^\mu s,
\eeq
where $f_s$ is an energy scale (i.e. conformal symmetry breaking scale for dilatons), $C_{s N}$ with $N=n$ or $p$ is a dimensionless coupling parameter, and again a factor of one half is conventional. The squared amplitude of the OPE diagrams for scalar neglecting $\eta$ meson exchange channels is found to be~\cite{Arndt:2002yg,Diener:2013xpa}:
\begin{align}
\sum_{\mathrm{spins}}|\mathcal{M}|=&8\left(\frac{C_{s N}}{f_s}\right)^2g_{\pi NN}^4\left[\left(\frac{\mathbf{k}^2}{\mathbf{k}^2+m_\pi^2}\right)^2+\left(\frac{\mathbf{l}^2}{\mathbf{l}^2+m_\pi^2}\right)^2+\frac{\mathbf{k}^2\mathbf{l}^2-2(\mathbf{k}\cdot\mathbf{l})^2}{(\mathbf{k}^2+m_\pi^2)(\mathbf{l}^2+m_\pi^2)}\right],\label{sME}
\end{align}
and again approximated as a constant:
\begin{align}
\sum_{\mathrm{spins}}|\mathcal{M}|\simeq&8\,g_{s NN}^2g_{\pi NN}^4,\label{sM}
\end{align}
where $g_{s NN}\equiv\frac{C_{sN}}{f_s}$.


These results indicate that although \eqref{sM} differs from \eqref{psM} by a factor of order unity, one does not need to worry about the difference given that uncertainties of supernova models and OPE approximation are present.

\subsubsection*{Phase Space Integration}
The phase space integration of the ALP energy production and absorption rate, \eqref{epdotint}-\eqref{lambdainvint}, for massless limit can be reduced to~\cite{Raffelt:2006cw,Raffelt:1996wa}:
\eq{
\dot{\epsilon}&=\xi\gN^2 \f{n_B}{16\pi^2}\int^{\infty}_0 dw\, w^4 S_{\sigma}(-w),\label{edotw}\\
\Gamma_w&=\xi\gN^2 \f{n_B}{8}w S_{\sigma}(w),\label{abspw}
}
where
\eq{
S_{\sigma}(w)\equiv \lim_{k\to0}S_{\sigma}(w,k)=\f{\Gamma_\s}{w^2+\Gamma_\s^2/4}s(w/T)\times
\begin{cases} 1  & w >0, 
\\ e^{w/T} &  w < 0,
\end{cases}\label{strfn}
}
is the long-wavelength approximation of the dynamical spin structure function where the momentum transfer to the medium is neglected, $\Gs$ is the spin fluctuation rate, and dimensionless integral $s(x)$ is approximated as $\sqrt{1+|x|\pi/4}$ within better than 2.2\% accuracy. I have also introduced a correction factor $\xi$ to compensate the approximations made in calculations, which will be discussed later.
The last conditional dependence on~\eqref{strfn} implies the quantum-mechanical detailed-balancing property,
\eq{
\dot{\epsilon}=\int\frac{d^3k}{(2\pi)^3}w\Gamma_w e^{-w/T}.
}
For ALP mass comparable to supernova temperature, its mass is no longer neglected and the absorption rate gets a factor $\sqrt{1-m_a^2/w^2}$ from the phase volume integration\footnote{Note that the phase space factor for massive ALP is multiplied to the massless limit results \eqref{edotw}-\eqref{abspw}. Although this sounds a naive approximation, I assert this would show an estimated tendency for very massive ALP as the constraints for an ALP mass comparable to the temperature already needs very precise knowledge of the supernova structures which still has many uncertainties. Numerical calculation of phase space integration for a massive ALP has been done in~\cite{LightRadion}.}. Defining $x\equiv w/T$, \eqref{edotw} then becomes
\eq{
\dot{\epsilon}&=\gamma\gN^2 \f{n_B\Gs T^3}{16\pi^2}\int^{\infty}_{m_a/T} dx\, \f{x^4}{x^2+\Gs^2/(2T)^2}\sqrt{1-\frac{m_a^2/T^2}{x^2}}s(x)e^{-x},
}
and the inverse mean free path of ALP having energy $w$ is
\eq{
\l^{-1}_x&=\Gamma_w\beta^{-1}=\gamma\gN^2 \f{n_B\Gs}{8T} \f{x^4}{x^2+\Gs^2/(2T)^2}\beta^{-1}s(x),
}
where $\beta=\sqrt{1-\frac{m_a^2}{w^2}}$.


Assuming local thermal equilibrium for the trapping regime, the absorption in \eqref{totaledot} at each point inside the star can be given by the reduced Rosseland mean free path,
\eq{
\l&=\frac{\intinfzero dw\,\l_w(1-e^{-w/T})^{-1}\beta_w \partial_T B_w(T)}{\intinfzero dw\,\beta_w \partial_T B_w(T)}\\
&= \left. \intinfzero dx\,\l_x\left(1-\frac{m_a^2/T^2}{x^2}\right)\f{x^4e^{2x}}{(1-e^{-x})^3} \middle/ \intinfzero dx\,\left(1-\frac{m_a^2/T^2}{x^2}\right)\f{x^4e^{x}}{(1-e^{-x})^2}\right.,\label{Rmfp}
}
where
\eq{
B_w(T)=\frac{1}{(2\pi^2)}\f{w^2(w^2-m_a^2)^{1/2}}{e^{w/T}-1}\label{bosonspectral}
}
is the boson spectral density for one spin degree of freedom.

It has been shown that the rate of soft pseudoscalar emission through bremsstrahlung ($NN\to NNa$) is directly related to the on-shell nucleon-nucleon scattering amplitude which can be extracted from experimentally measured nucleon-nucleon scattering phase-shift data~\cite{Hanhart:2000ae}. As such a calculation doesn't need OPE approximation, it has an advantage that it is model-independent. It has been shown that the OPE is an overestimate by about a factor of three compared with the model-independent calculation. Meanwhile, such a model-independent technique does not work for scalar emission ($NN\to NNs$) due to a cancellation of the leading infrared pole term in the sum which renders low energy theorems unable to be constructed~\cite{Arndt:2002yg}. In such a case the coupling of a scalar to unknown strong interaction vertices is no longer subleading, and the model-independent calculation cannot be achieved. However, I'll assume that OPE approximation for scalar emission also overestimates reality by a similar amount and has the same correction factor $\xi=1/3$.

%
%
%
%
%
%

\subsection{Primakoff Process}
\begin{figure}[thpb]
 \centering
  \includegraphics[height=.18\textheight]{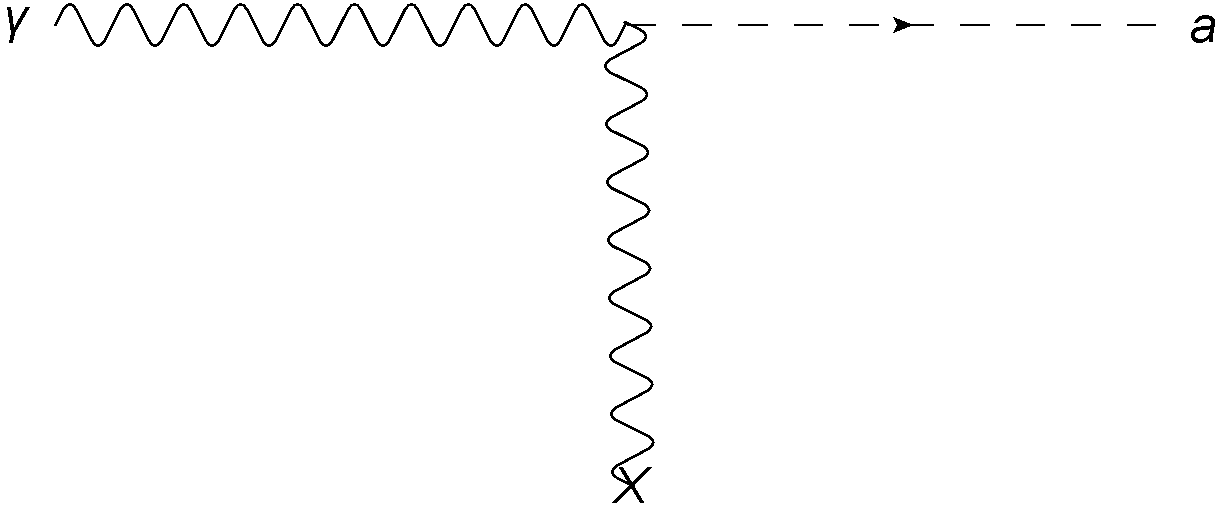}
  \caption{Primakoff Process in an external electromagnetic field through the two photon coupling of ALP.}
  \label{fig:Primakoff}
\end{figure}

The two photon coupling of ALPs,
\beq
\mathcal{L}_{a\gamma\gamma}=-\frac{1}{4}g_{a\gmgm}F_{\mu\nu}\tilde{F}^{\mu\nu}a=g_{a\gmgm}{\bf{E\cdot B}}\,a,
\eeq
for pseudoscalar, and
\beq
\mathcal{L}_{s\gamma\gamma}=-\frac{1}{4}g_{s\gmgm}F_{\mu\nu}F^{\mu\nu}s=g_{s\gmgm}\frac{1}{2}({\bf{E^2-B^2}})s,
\eeq
for scalar allow the production and absoption channel primarily through Primakoff process~\cite{Raffelt:1996wa} shown in Fig.~\ref{fig:Primakoff}.

The transition rate for pseudoscalar is calculated as~\cite{Raffelt:1996wa}
\eq{
\Gamma_{\gamma\to a}=\frac{g_{a\gmgm}^2 T k_S^2}{32\pi}\left[\left(1+\frac{k_s^2}{4w^2}\right)\ln\left(1+\frac{4w^2}{k_s^2}\right)-1\right],\label{PrimaRate}
}
where the plasma effective mass of the initial state photons and the ALP mass are neglected relative to the energy $w$, and the forward divergence from the long-range Coulomb interaction is cut off in a plasma by the inverse screening length~\cite{Raffelt:1985nk},
\eq{
k_s^2\equiv \frac{4 \pi \alpha }{T}\left(n_e+\sum_i Z_i^2 n_i \right),
}
where $i$ represents various nuclear species, $\alpha$ is the electromagnetic fine-structure constant.
I assume the SN core is neutron-rich and $n_p/n_n=0.1$. Given that nuclei in SN core are partially degenerate, the effective number of targets would be reduced. The effective number of targets, $n^{eff}_i$, varies from 0.2 $n_i$ to 0.8 $n_i$~\cite{Payez:2014xsa} in the SN core, but for simplicity I assume $n^{eff}_i=n_i/2$. Under the SN core conditions, electrons are highly degenerate and their phase space is Pauli-blocked, hence give negligible contribution. Thus one can evaluate \eqref{PrimaRate} using the effective inverse screening length,
\eq{
k_s^2 \to \frac{4 \pi \alpha }{T} Z_{i}^2  n^{eff}_{i}.
}
 The explicit calculation~\cite{Marsh:2014gca} shows \order{1} factor difference between a pseudoscalar and scalar transition rate. As discussed in \ref{Bremsstrahlung}, these difference can be neglected.


\subsubsection*{Phase Space Integration}\label{ElossPrima}

In the same manner as discussed in \ref{Bremsstrahlung}, the massive ALP production is obtained as
\eq{
\dot{\epsilon}&=\int\frac{2d^3k}{(2\pi)^3}\f{w\Gamma_{\gamma\to a}}{e^{w/T}-1}\sqrt{1-\frac{m_a^2/T^2}{x^2}}=\int^{\infty}_{m_a/T}\frac{dx}{\pi^2}\f{x^3 \,T^4\,\Gamma_{\gamma\to a} }{e^{x}-1}\sqrt{1-\frac{m_a^2/T^2}{x^2}}\label{Primaedot},
}
where two polarizations of initial state photons, a Bose stimulation factor and massive ALP phase volume factor are explicitly included. The inverse mean free path depends both on inverse Primakoff process and two photon decays. The decay rate to two photons is
\eq{
\Gamma_{\mathrm{decay}}=\frac{g_{a\gmgm}^2 m_a^3}{64\pi}.
}
Comparing with \eqref{PrimaRate}, $\Gamma_{\mathrm{decay}}$ won't affect the result unless $m_a^3 \sim 8\pi n_{p}$ or larger, which also saturates $m_a \sim 10T$ or larger.
%
Given mass and kinetic energy, one has
\eq{
\lambda_x^{-1}&=\Gamma_{a\to\gamma}\beta^{-1}+\Gamma_{\mathrm{decay}}(\gamma\beta)^{-1}\label{Primamfp}.
}
where 
$\gamma$ is a Lorentz factor and $\Gamma_{a\to\gamma}= n^{eff}_{p}(\s v)_{a\to\gamma}$ with $\s _{a\to\gamma}=\frac{2}{\beta}\s _{\gamma\to a}$~\cite{Dolan:2017osp}. The reduced Rosseland mean free path is obtained by plugging~\eqref{Primamfp} into \eqref{Rmfp}.

%
%

%
%

%
%
%
%
%
%
%
%

\section{Results}\label{Chap:Results}
\begin{figure}[thpb]
 \centering
  \includegraphics[height=.4\textheight]{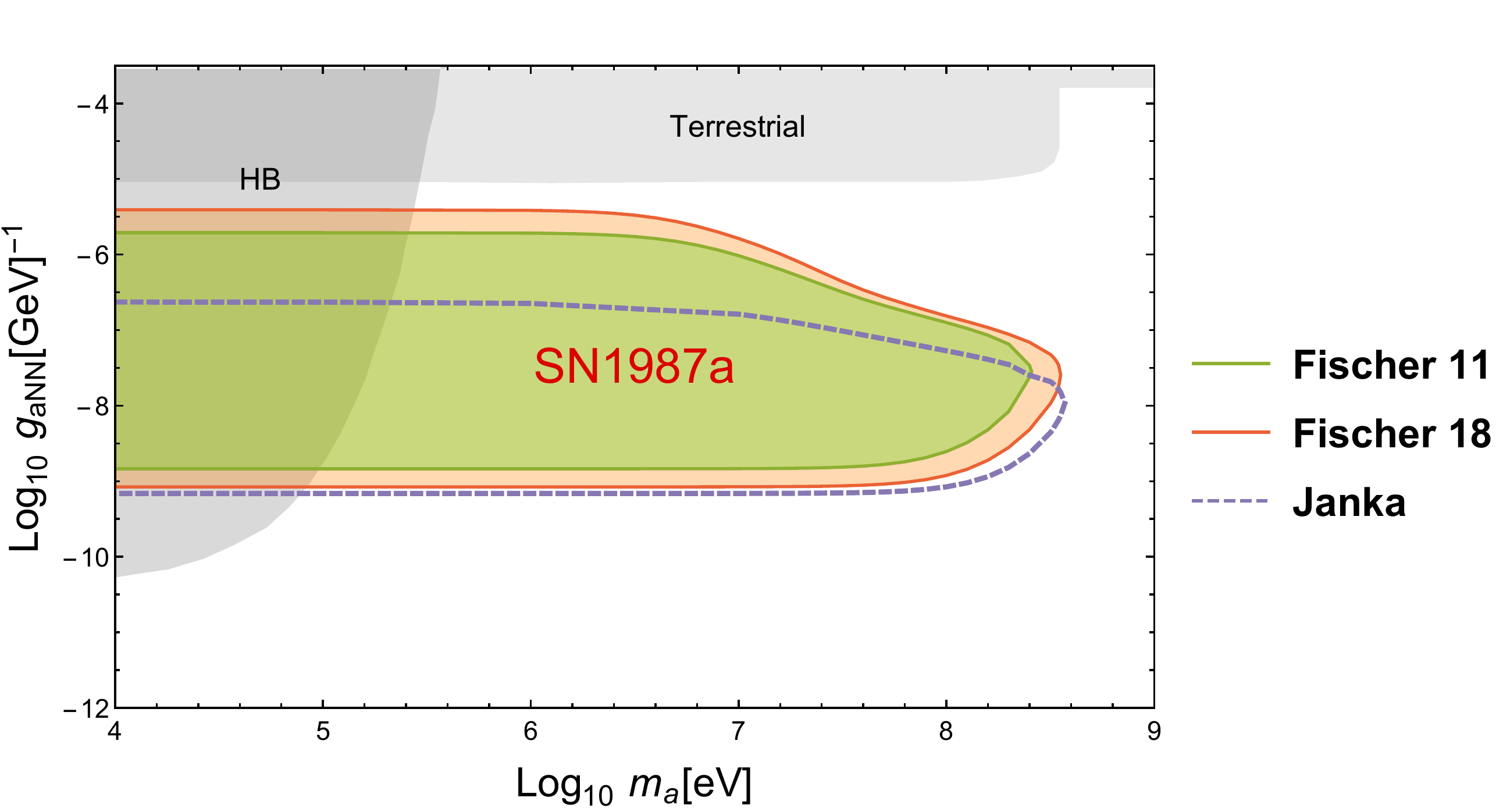}
  \caption{Exclusion plot from the nucleon-nucleon Bremssrahlung through the two-nucleon coupling. Cooling limits from HB stars \cite{Hardy:2016kme} are shown in gray, and terrestrial constraints that are model-dependent are shown in light-gray for $g_{aNN}\simeq 0.3\,g_{att}$~\cite{Knapen:2017xzo} (see text).}
  \label{fig:BremConstraints}
\end{figure}

\begin{figure}[thpb]
 \centering
  \includegraphics[height=.4\textheight]{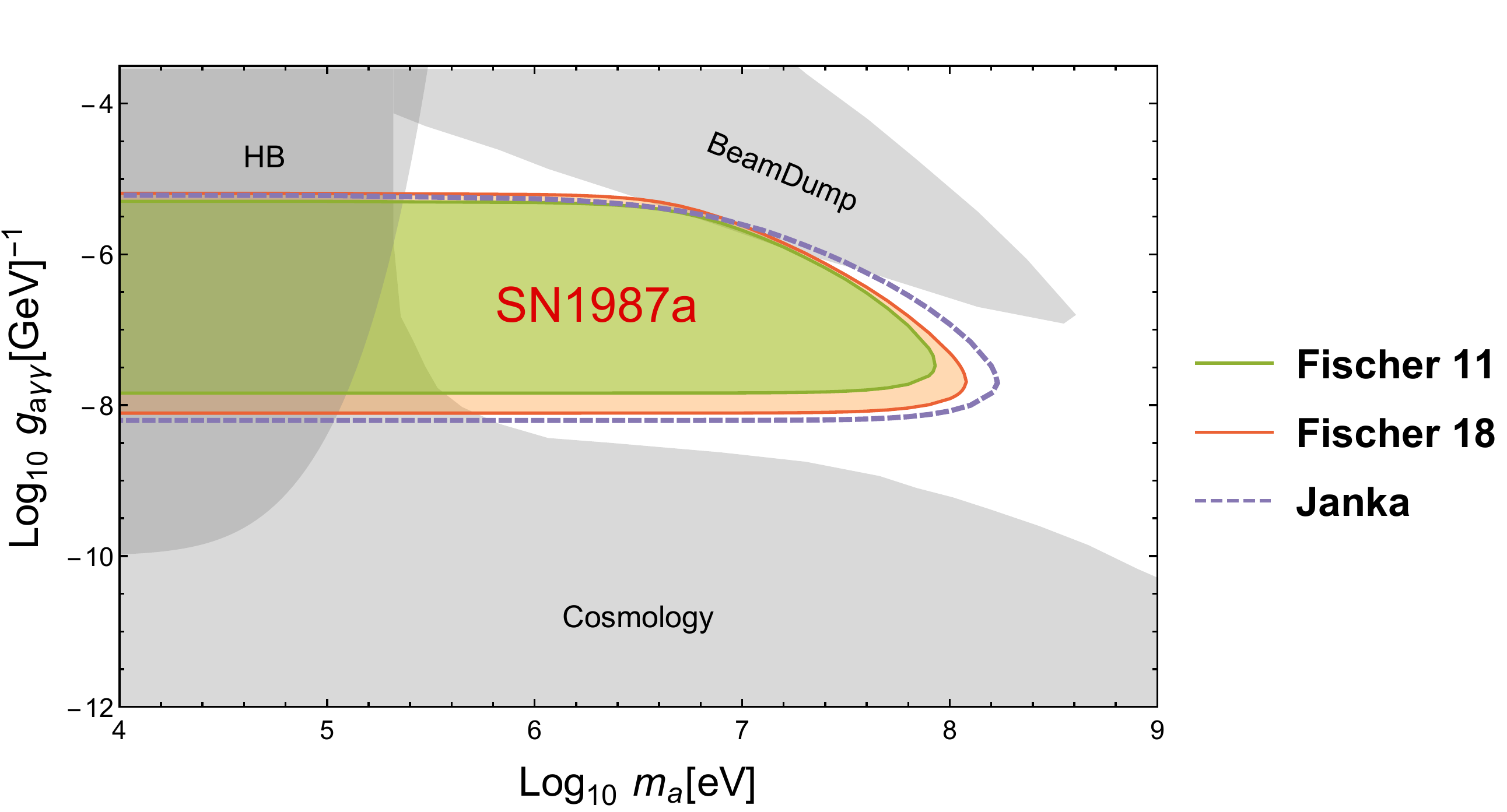}
  \caption{Exclusion plot from the Primakoff Process in an external electromagnetic field in SN1987a through the two-photon coupling of ALP. Other constraints in gray are compiled from Fig. \ref{fig:ALPConstraints}.}
  \label{fig:PrimakoffConstraints}
\end{figure}

The ALP constraints by SN1987a in mass-coupling parameter space is plotted in Fig.~\ref{fig:BremConstraints} and Fig.~\ref{fig:PrimakoffConstraints}. In Fig.~\ref{fig:BremConstraints}, the excluded range by \eqref{totEbound} is shown for the ALP coupling to nucleons when ALPs are produced and absorbed through nucleon-nucleon bremsstrahlung when other couplings are all neglected. Stellar cooling limits from HB stars \cite{Hardy:2016kme}, and terrestrial constraints~\cite{Knapen:2017xzo} are also shown in Fig.~\ref{fig:BremConstraints}. The terrestrial constraints from exotic meson decays, i.e. $B\to K a$ and $K\to \pi a$, are model-dependent and in particular sensitive to a coupling to top\footnote{A coupling to top is defined as $\mathcal{L}\subset g_{att}m_t a t\bar{t}$}. $g_{aNN}\simeq 0.3\,g_{att}$ is assumed for the terrestrial constraints in Fig.~\ref{fig:BremConstraints}, which is one of the scenarios in \cite{Knapen:2017xzo}. In Fig.~\ref{fig:PrimakoffConstraints}, the excluded range by \eqref{totEbound} is shown for the ALP coupling to photons when ALPs are produced and absorbed through Primakoff process when other couplings are all neglected.

\begin{figure}[thpb]
	\centering
	\includegraphics[height=.4\textheight]{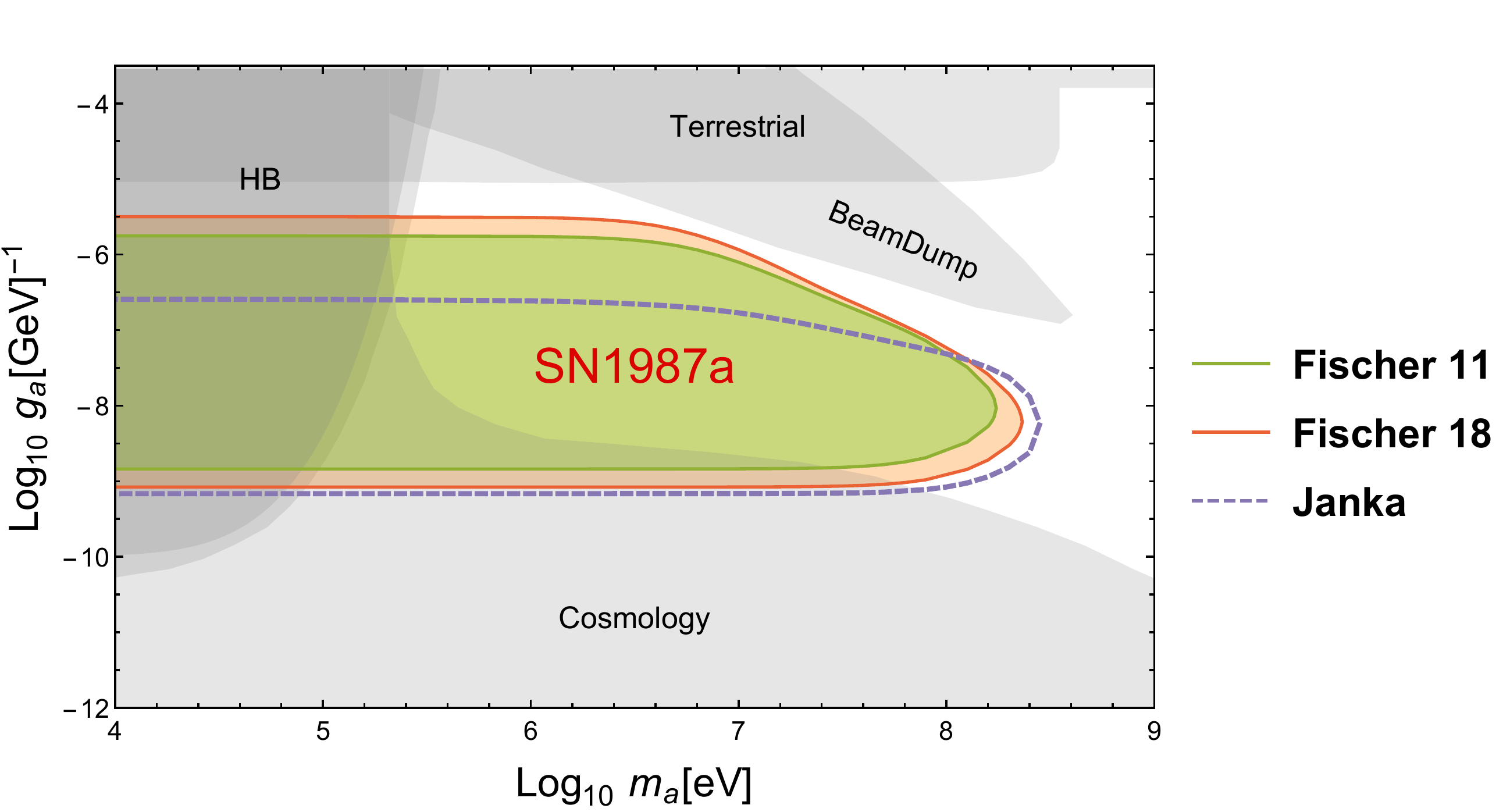}
	\caption{Exclusion plot from both the nucleon-nucleon Bremssrahlung through the two-nucleon coupling and the Primakoff Process through the two-photon coupling with $g_a\equiv g_{aNN}=g_{a\gamma\gamma}$. Other constraints in gray are copied from Fig. \ref{fig:BremConstraints} and Fig. \ref{fig:PrimakoffConstraints}.}
	\label{fig:BremPrimaConstraints}
\end{figure}

In Fig.~\ref{fig:BremPrimaConstraints}, I provide the excluded range by \eqref{totEbound} for the case $g_{aNN}=g_{a\gamma\gamma}(\equiv g_a)$ where ALPs are produced and absorbed through both the nucleon-nucleon bremsstrahlung and the Primakoff process.

While preparing this work, similar work on SN1987a constraints to ALP-nucleon coupling has been done in~\cite{Chang:2018rso}, and it is worthwhile to compare different parameters and assumptions which cause a little different result. First, while other correction factors stated in \cite{Chang:2018rso} are similarly accounted, for the correction for spin structure function, $S_\s/(S_\s|_{\mathrm{OPE}})$ denoted as $\gamma_h$ in \cite{Chang:2018rso}, they refer the result shown in \cite{Bartl:2016iok}. However, the temperature and density function used in \cite{Bartl:2016iok} are not exactly compatible with profiles from Fischer et al.~\cite{Fischer} which I and authors of \cite{Chang:2018rso} use. In this work, a uniform correction $\xi=1/3$ is approximately used as explained in Sec.~\ref{Bremsstrahlung}. Applying $\gamma_h$ as shown in \cite{Chang:2018rso} shifts the bound on $g_{aNN}$ in Fig.~\ref{fig:BremConstraints} down by a factor of unity. Secondly, while the density and temperature in the very inner core, $r<5$km, are not provided in~\cite{Fischer}, they are assumed to be constant in this work, and interpolation functions~\cite{privateconv} are used in \cite{Chang:2018rso}. The free streaming limits are affected with this difference; however, it does not comparably amend the trapping limits in that the profiles of the outer core or mantle are more important for trapping limits. Note that a $O(1)$ factor correction on \eqref{epdotint} and \eqref{lambdainvint} results in $\sqrt{O(1)}$ correction to the couplings. Together with uncertainties of the supernova structure, we will not worry about a factor of unity for the present discussion.



\section{Conclusions}\label{Chap:Conclusions}
The constraints on a scalar/pseudoscalar Axion-Like-Particle imposed by a criterion \eqref{totEbound} from neutrino detections made right after Supernova 1987a are revisited. As shown in \cite{LightRadion}, a dilaton/radion may have far different coupling strength to photons and to nucleons, which results in the fact that each limit can be alleviated with the aid of the other. Although there isn't a known natural mechanism to have very different coupling to photons and nucleons (or any fermions) for a pseudoscalar Axion-like-Particle, it is worth looking in detail at the constraints enforced on each interaction independently for a better understanding of the exclusion regions in the ALP parameter space.
I have found that broader range of $g_{aNN}$ is excluded by SN1987a relative to $g_{a\gamma\gamma}$ limits in Fig.~\ref{fig:ALPConstraints} where it is assumed $g_{aNN}\sim g_{a\gamma\gamma}$. From the Primakoff process, the excluded range of $g_{a\gamma\gamma}$ is directly attained, which differs by an order of magnitude compared with Fig.~\ref{fig:ALPConstraints}. For an axion-like case in Fig.~\ref{fig:BremPrimaConstraints}, i.e. $g_{aNN}=g_{a\gamma\gamma}$, I have found the trapping regime for $m_a\gtrsim10\MeV$ is alleviated compared to previous bounds. These new open windows need to be probed by future experiments.  


\section*{Acknowledgments}
JL thanks John Terning, Thomas Flacke, Jae Hyeok Chang, and Samuel D. McDermott for useful discussions; thanks Joerg Jaeckel, and Michael Spannowsky for providing their data and plots.



%
%
%
%
%
%

%
%

\end{document}